\documentclass[a4paper]{jpconf}
\usepackage{graphicx}
\usepackage{amsmath}
\usepackage{amssymb}
\usepackage{lineno}
\usepackage{cite}
\usepackage{wrapfig}
\usepackage{epstopdf}
\begin{document}
\title{Measurement of D-meson production in pp, p--Pb, and Pb--Pb collisions at the LHC with the ALICE detector}

\author{Chitrasen Jena, for the ALICE Collaboration}

\address{Universit\`a  degli Studi di Padova and INFN Sezione di Padova}

\ead{jena@pd.infn.it}

\begin{abstract}Heavy quarks are a powerful probe for investigating the properties of the Quark-Gluon Plasma created in heavy-ion collisions, since 
they are produced in initial hard scattering processes and experience all the stages of the medium evolution. ALICE has measured the production of 
$\mathrm{D}^0$, $\mathrm{D}^+$, $\mathrm{D}^{*+}$, and $\mathrm{D}_{\mathrm{s}}^+$ mesons at central rapidity in their hadronic decay channels 
in various collision systems and energies. We present recent results for D-meson production measured by the ALICE Collaboration in pp collisions 
at $\sqrt{s}=7$ and 2.76~TeV, Pb--Pb collisions at $\sqrt{s_\mathrm{NN}}=2.76$~TeV and p--Pb collisions at $\sqrt{s_\mathrm{NN}}=5.02$~TeV.
\end{abstract}

\section{Introduction}\label{sec:intro}
The measurement of heavy-flavour production in various collision systems and energies offers the opportunity to investigate the properties 
of the hot and dense QCD matter, the Quark-Gluon Plasma (QGP), created in high energy heavy-ion collisions. Due to their large masses, 
charm and beauty quarks are produced predominantly in hard parton scattering processes in the early stage of the collision. 
They interact with the medium created in the collision and lose energy via both inelastic (medium-induced gluon radiation, or radiative energy loss)~\cite{ineloss1,ineloss2}
and elastic (collisional energy loss)~\cite{eloss1,eloss2} processes. Theoretical calculations for energy loss predict a dependence on the colour charge and mass 
of the parton propagating through the medium. In QCD, quarks have a smaller colour coupling factor with respect to gluons, so that the energy loss for 
quarks is expected to be smaller than for gluons. The ‘dead-cone effect’ should reduce small-angle gluon radiation for heavy quarks with 
moderate energy-over-mass values and hence the radiative energy loss for heavy quarks is expected to be smaller than for light ones~\cite{deadcone}. 
In addition, collisional energy loss is expected to be reduced for heavier quarks, because the spatial diffusion coefficient that regulates the momentum transfer 
with the constituents of the medium is expected to scale as $1/m_{q}$~\cite{collEloss}. 

A sensitive observable to characterize the effect of the dense medium on the measured heavy-flavour hadron yields is the nuclear modification factor, defined as
\begin{equation}
R_\mathrm{AA}(p_\mathrm{T}) = \frac{1}{\left<T_\mathrm{AA}\right>}\frac{\mathrm{d}N_\mathrm{AA}/\mathrm{d}p_\mathrm{T}}{\mathrm{d}\sigma_\mathrm{pp}/\mathrm{d}p_\mathrm{T}}, 
\label{RAA}
\end{equation}
where $\left<T_\mathrm{AA}\right>$ is the average nuclear overlap function estimated using the Glauber model~\cite{glauber}, $\mathrm{d}N_\mathrm{AA}/\mathrm{d}p_\mathrm{T}$ 
is the transverse momentum differential yield measured in nucleus-nucleus collisions and $\mathrm{d}\sigma_\mathrm{pp}/\mathrm{d}p_\mathrm{T}$ is the $p_\mathrm{T}$-differential 
production cross section in pp collisions. In-medium energy loss results in a suppression, $R_\mathrm{AA}<1$, of hadron yields 
at moderate to high transverse momentum. The dependence of the energy loss on the parton species and mass can be investigated by 
comparing the nuclear modification factors of hadrons with charm $(R_\mathrm{AA}^{\mathrm{D}})$ and beauty $(R_\mathrm{AA}^{\mathrm{B}})$ with that of pions $(R_\mathrm{AA}^{\pi})$,
mostly originating from gluon fragmentation at LHC energies. The expected dependence of the energy loss on the parton species should determine a difference in their $R_\mathrm{AA}$. 
However, it is important to note that the comparison of heavy-flavour hadron and pion $R_\mathrm{AA}$ cannot be interpreted directly as a comparison of charm, beauty, and gluon 
energy losses, due to the different parton fragmentation functions and slopes of the $p_\mathrm{T}$-differential cross section. For example, the theoretical calculations in 
Ref.~\cite{ElosspiD} shows similar $R_\mathrm{AA}$ for charmed hadrons and pions despite the different energy loss of c quarks, light quarks and gluons. Moreover, at 
low $p_\mathrm{T}$, a significant fraction of pions are produced from soft processes, not hard-scatterings.

The measurement of anisotropy in the azimuthal distribution of particle momenta provides further insight into the properties of the QGP. In heavy-ion collisions, 
anisotropic patterns of particle production originate from the initial anisotropy in the spatial distribution of the nucleons participating in the collision. The azimuthal 
anisotropy of produced particles is characterized by the Fourier coefficients $v_{n}=\langle \cos(n(\varphi-\Psi_\mathrm{RP})) \rangle$, where $n$ is the order of the
harmonic, $\varphi$ the azimuthal angle of the particle and $\Psi_\mathrm{RP}$ is the azimuthal angle of the reaction plane. The reaction plane is spanned by the vector 
of the impact parameter and the beam direction. For non-central collisions, the overlap region of the colliding nuclei has a lenticular shape and the azimuthal anisotropy 
is dominated by the second Fourier coefficient $v_2$, commonly denoted elliptic flow. At low and intermediate $p_\mathrm{T}$, the $v_2$ of heavy-flavour hadrons is 
sensitive to the degree of thermalisation of charm quarks in the medium~\cite{lowpTv2}, while at higher $p_\mathrm{T}$ it allows the study of the path-length dependence of 
in-medium parton energy loss~\cite{highpTv2}.

The measurement of charm and beauty production cross sections in pp collisions at the LHC provides an important test of perturbative QCD (pQCD) 
calculations and serves as an essential reference for understanding the results from heavy-ion collisions~\cite{alicepbpbraa}. In addition, heavy-flavour production 
as a function of charged-particle multiplicity in pp collisions could provide insight into the relevance of multiple hard scattering processes. Initial-state effects, such as 
the modification of parton distribution functions in the nucleus and momentum broadening due to parton scattering in the nucleus, affect heavy-quark production 
and cannot be accounted for with a reference from pp data.  It is therefore 
necessary to study p--Pb collisions, where an extended hot and dense strongly-interacting medium is not expected to form, to quantify these initial-state effects.

In these proceedings, we will discuss the ALICE measurements of D-meson production in pp collisions at $\sqrt{s}=7$~TeV, Pb--Pb collisions 
at $\sqrt{s_\mathrm{NN}}=2.76$~TeV, and p--Pb collisions at $\sqrt{s_\mathrm{NN}}=5.02$~TeV. 

\section{D-meson decay reconstruction with ALICE}
The D-meson decays are reconstructed in the central rapidity region utilizing the tracking and particle identification capabilities of the ALICE central barrel detectors. 
The central barrel detectors are surrounded by a large solenoidal magnet, which provides a magnetic field of 0.5 T along the beam direction. A detailed description 
of the ALICE apparatus is given in Ref.~\cite{ALICEdet}.

The measurement of charm production is performed by reconstructing $\mathrm{D}^0$, $\mathrm{D}^+$, $\mathrm{D}^{*+}$ and $\mathrm{D}_{\mathrm{s}}^+$ 
charmed hadrons via their hadronic decays 
$\mathrm{D}^0\rightarrow{}\mathrm{K}^{-}\pi^+ ($with branching ratio, BR, of $3.88 \pm 0.05\%)$, 
$\mathrm{D}^+\rightarrow{}\mathrm{K}^{-}\pi^+\pi^+($BR of $9.13 \pm 0.19\%)$,
$\mathrm{D}^{*+}\rightarrow{}\mathrm{D}^0\pi^+ ($BR of $67.7 \pm 0.5\%)$ with $\mathrm{D}^0\rightarrow \mathrm{K}^-\pi^+$ and
$\mathrm{D}_{\mathrm{s}}^+\rightarrow\phi\pi^+\rightarrow\mathrm{K}^+\mathrm{K}^-\pi^+ ($BR of $2.28 \pm 0.12\%)$ 
and their charge conjugates~\cite{PDGreview}. The $\mathrm{D}^0$, $\mathrm{D}^+$, and $\mathrm{D}_{\mathrm{s}}^+$ mesons have mean proper decay lengths 
$c\tau \approx$ 123, 312, and 150 $\mu\mathrm{m}$, respectively~\cite{PDGreview}. Therefore, their secondary decay vertices are typically displaced by a 
few hundred $\mu\mathrm{m}$ from the primary interaction vertex. The analysis strategy for the $\mathrm{D}^0$, $\mathrm{D}^+$, and 
$\mathrm{D}_{\mathrm{s}}^+$ is based on the reconstruction and selection of secondary vertex topologies with significant separation from
the primary vertex. In the case of the $\mathrm{D}^{*+}$, which decays strongly, the secondary vertex topology of the daughter $\mathrm{D}^0$ is reconstructed. 
The topological reconstruction of the decays allows for an efficient rejection of the combinatorial background from 
uncorrelated tracks. The Time Projection Chamber (TPC) and the Inner Tracking System (ITS) detectors, in particular the two innermost ITS layers 
equipped with Silicon Pixel Detectors (SPD), provide a resolution of the track impact parameter of about 60 $\mu\mathrm{m}$ for tracks with 
$p_\mathrm{T} = 1$ GeV/$c$ in Pb--Pb collisions. In order to further suppress the combinatorial background, measurements of the particle time-of-flight 
from the collision point to the Time Of Flight (TOF) detector and the specific energy loss in the TPC gas are used to identify kaons and pions. 
Finally, the signal yield is extracted by fitting the invariant mass distribution with a Gaussian function for the signal and an exponential 
function for the background shape for $\mathrm{D}^0$, $\mathrm{D}^+$, and $\mathrm{D}_{\mathrm{s}}^+$. The $\mathrm{D}^{*+}$ signal is 
calculated by fitting the invariant-mass difference $\Delta m=m_{\rm D^{*+}}-m_{\rm D^0}$ with a function that consists of a Gaussian term describing 
the signal and the term  $a\,\sqrt{\Delta m-m_\pi}\cdot {\rm e}^{b(\Delta m-m_\pi)}$ for the background, where $m_{\pi}$ is the charged-pion mass, and 
$a$ and $b$ are free parameters. The raw yields extracted with the fit are corrected for acceptance and efficiency using Monte Carlo simulations based 
on Pythia~\cite{Pythia}  and HIJING event generators~\cite{Hijing}.

In order to extract the cross section of prompt D mesons, the contribution of D mesons from B decays is subtracted. This B feed-down contribution 
is evaluated using the B-meson cross section from the FONLL pQCD calculations~\cite{FONLL}, the $\mathrm{B}\rightarrow{}\mathrm{D}+\mathrm{X}$ decay 
kinematics from the EvtGen package~\cite{EvtGen}, the efficiencies for prompt and feed-down D mesons from simulations, and a hypothesis on 
the $R_\mathrm{AA}$ ($R_\mathrm{pPb}$) of D mesons arising from B-meson decays in Pb--Pb (p--Pb) collisions. Details of the analysis procedure
in pp, Pb--Pb, and p--Pb collisions are reported in Refs.~\cite{ALICEpp7TeVDmesons},~\cite{alicepbpbraa}, and~\cite{aliceRpPb}, respectively.

\section{Results in pp collisions} \label{sec:ppresults}
\begin{figure}[h!t]
\centering
\begin{minipage}[t]{13.9pc}\vspace{0px}
\includegraphics[width=13.9pc]{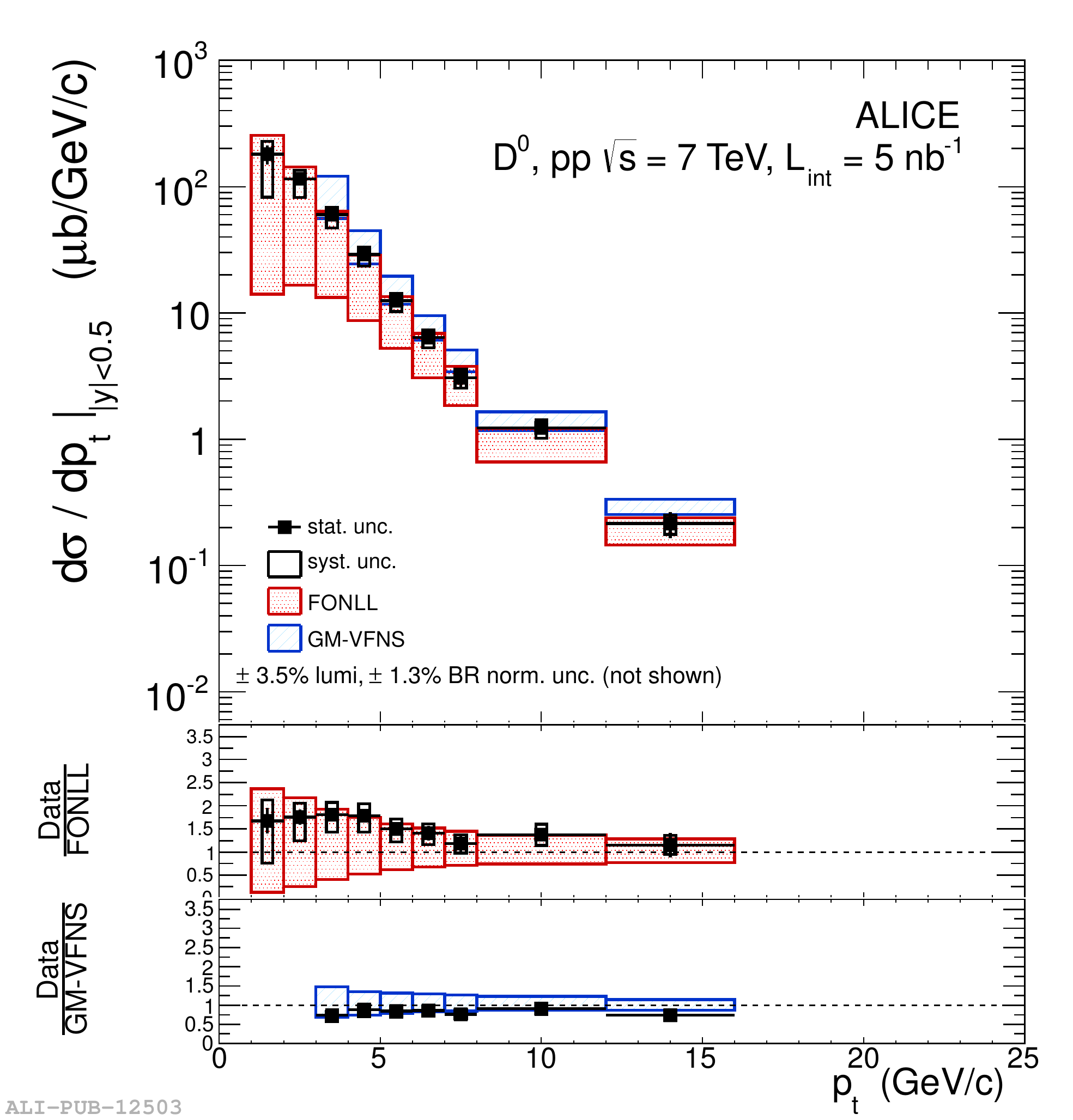}
\caption{\label{fig:diffcrosssectionspp7TeV}$p_\mathrm{T}$-differential cross section of prompt D$^{0}$ in pp collisions at $\sqrt{s}=7$~TeV, compared 
with results from FONLL~\cite{FONLL} and GM-VFNS~\cite{GM-VFNS} theoretical calculations~\cite{ALICEpp7TeVDmesons}.}
\end{minipage}\hspace{2pc}%
\begin{minipage}[t]{18.3pc}\vspace{-1.8px}
\includegraphics[width=18.3pc]{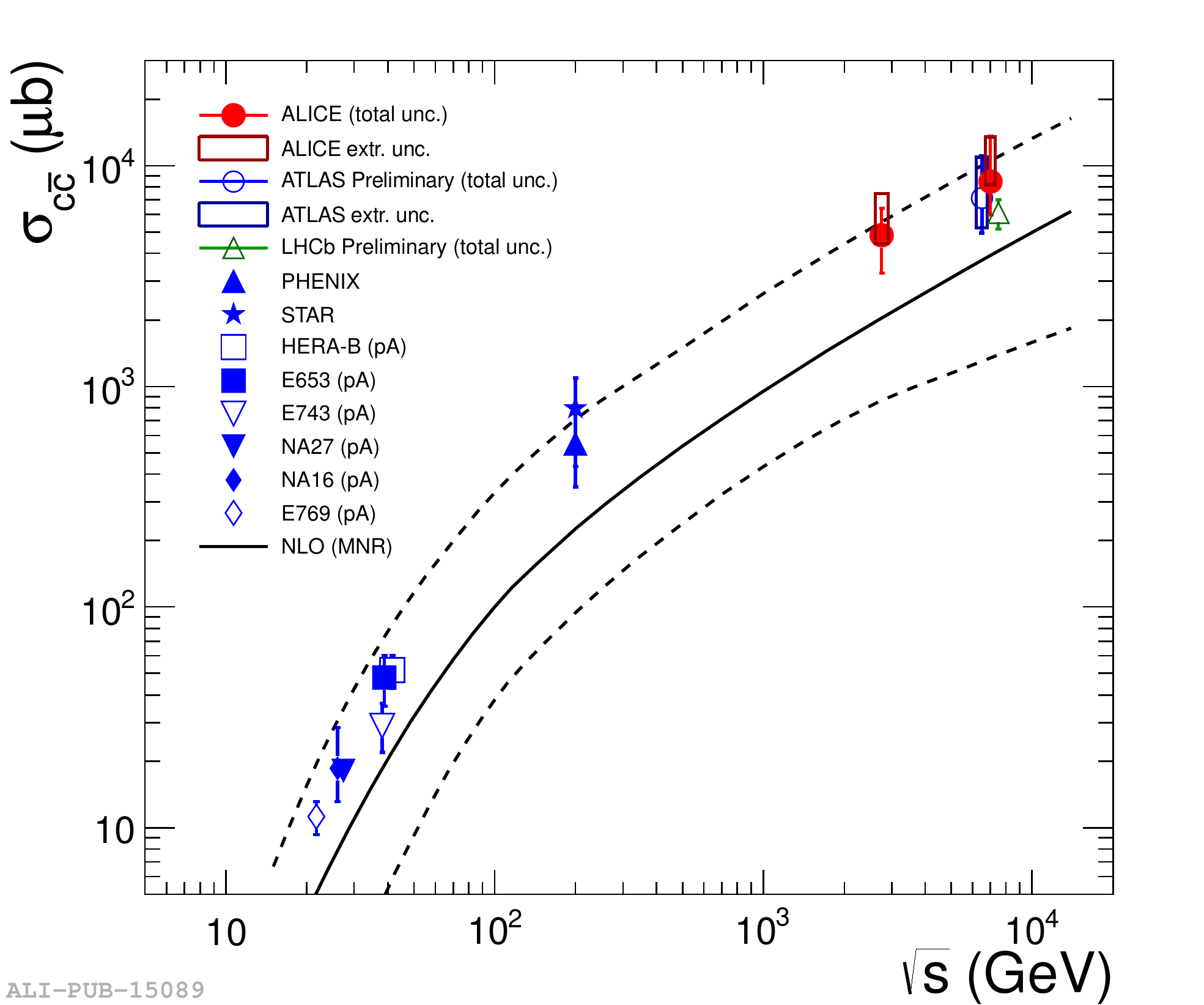}\vspace{-7px}
\caption{\label{fig:totalcharmcrosssectionpp}Total charm cross section measured in pp collisions with ALICE as a function of $\sqrt{s}$, in comparison 
with other experimental results and MNR~\cite{MNR} calculations~\cite{ALICEpp276TeVDmesons}.}
\end{minipage} 
\end{figure}
The measurement of prompt D-meson production in pp collisions at $\sqrt{s}=7$~TeV was performed using a data sample collected in 2010 with a minimum-bias trigger. 
The analysed data correspond to an integrated luminosity of 5~nb$^{-1}$. The measured $p_\mathrm{T}$-differential cross section for prompt $\mathrm{D}^0$, 
$\mathrm{D}^+$, $\mathrm{D}^{*+}$, and $\mathrm{D}_{\mathrm{s}}^+$ mesons are published in Ref.~\cite{ALICEpp7TeVDmesons,ALICEpp7TeVDsmesons}. 
Figure~\ref{fig:diffcrosssectionspp7TeV} shows the differential cross section of prompt D$^{0}$ production in the $p_\mathrm{T}$ range $1 < p_\mathrm{T} < 16$~GeV/$c$. The 
measured cross sections are described within uncertainties by pQCD calculations such as FONLL~\cite{FONLL},  GM-VFNS~\cite{GM-VFNS} and 
$k_\mathrm{T}$-factorization at LO~\cite{kTfact}. The data are at the upper edge of the uncertainty band for FONLL and $k_\mathrm{T}$-factorization model, but 
at the lower edge for GM-VFNS. 

The D-meson production cross section has also been measured in pp collisions at $\sqrt{s}=2.76$~TeV~\cite{ALICEpp276TeVDmesons}. This 
data sample was collected in 2011 with an integrated luminosity of 1.1~nb$^{-1}$, much smaller than that recorded at $\sqrt{s}=7$~TeV. 
Therefore, these data were not used  as the pp reference for the Pb--Pb results, but only to verify the validity of an energy scaling of 
the cross section from centre-of-mass energy 7~TeV to 2.76~TeV using FONLL~\cite{energyscaling}.

The measured cross sections for pp collisions at $\sqrt{s}=7$~TeV and 2.76~TeV were extrapolated to full phase space using FONLL calculations, in order to extract the total 
c$\bar{\mathrm{c}}$ cross section in pp collisions at these energies. The extrapolated results are shown in Fig.~\ref{fig:totalcharmcrosssectionpp}, along with 
the results from other experiments and a next-to-leading-order pQCD calculation (MNR~\cite{MNR}). The ALICE result at $\sqrt{s}=7$~TeV is in 
agreement with those from the ATLAS~\cite{ATLASpp} and LHCb~\cite{LHCbpp} Collaborations, and both of the ALICE points follow the trends exhibited by the 
NLO predictions and other experimental results spanning three orders of magnitude in energy.
\begin{figure}[!htb]
\centering
\begin{minipage}{12.1pc}\vspace{-0.0px}
\includegraphics[width=12.8pc]{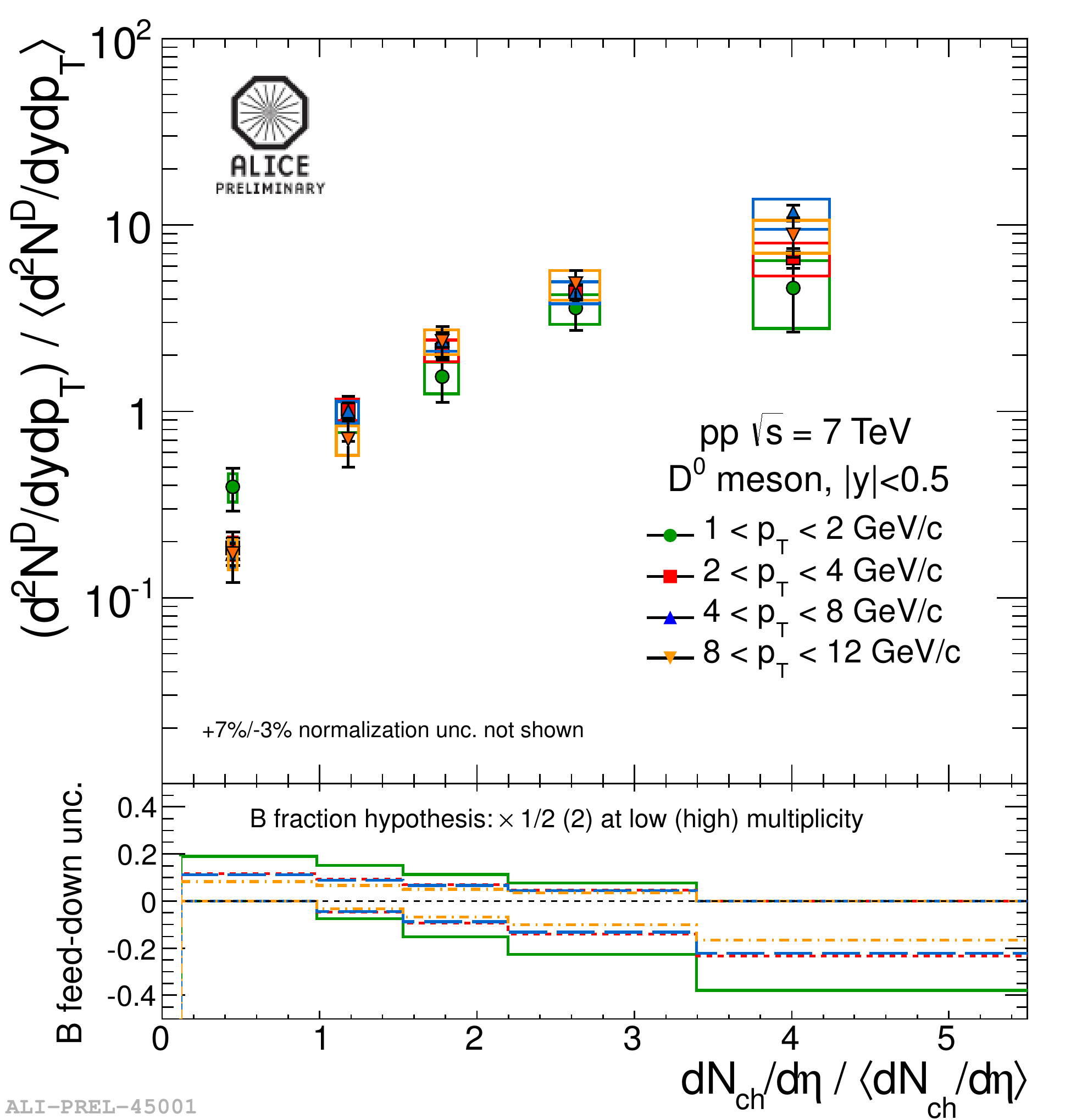}
\end{minipage}
\begin{minipage}{12.1pc}\vspace{-0.0px}
\includegraphics[width=12.8pc]{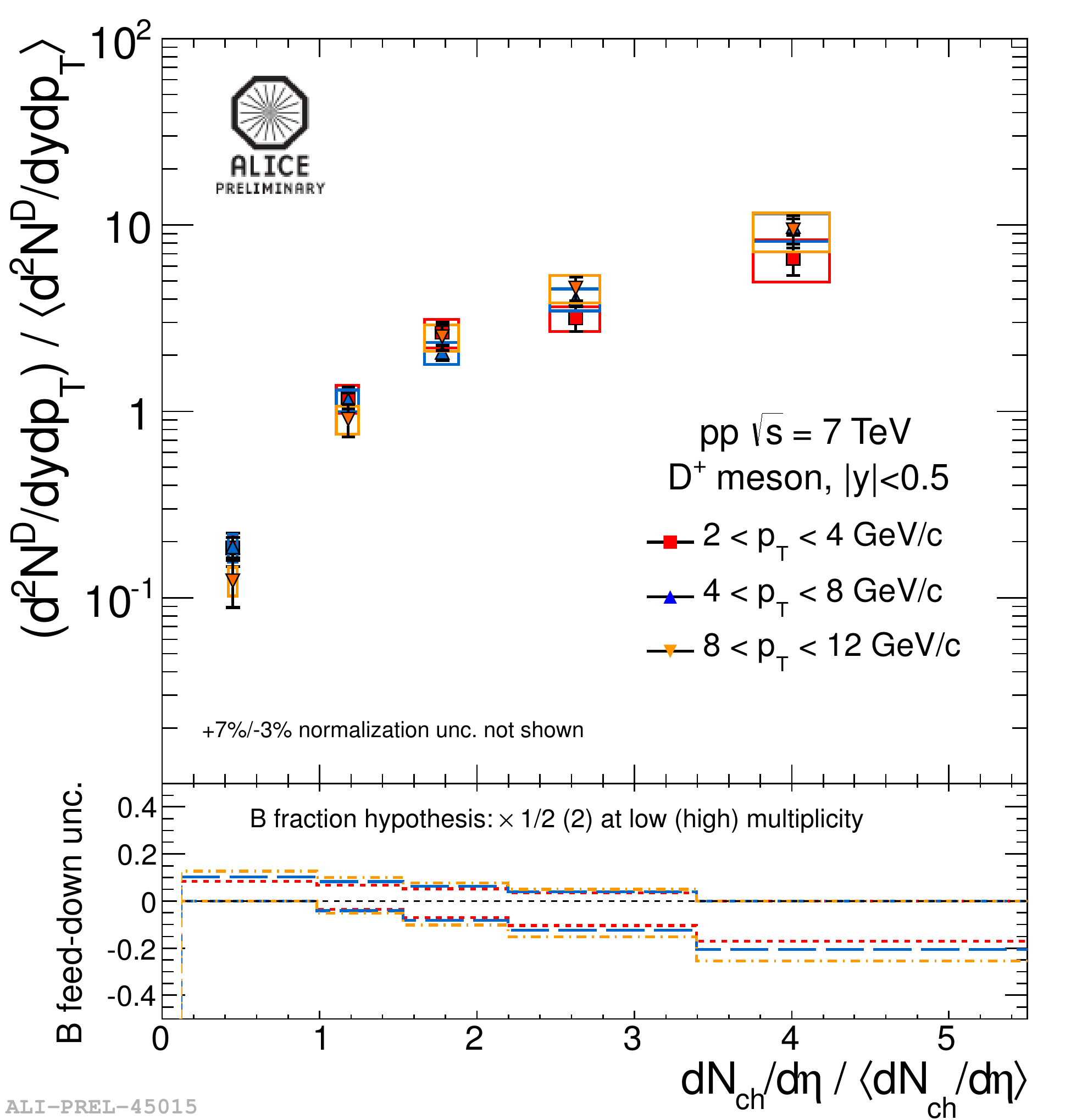}
\end{minipage} 
\begin{minipage}{12.1pc}\vspace{-0.0px}
\includegraphics[width=12.8pc]{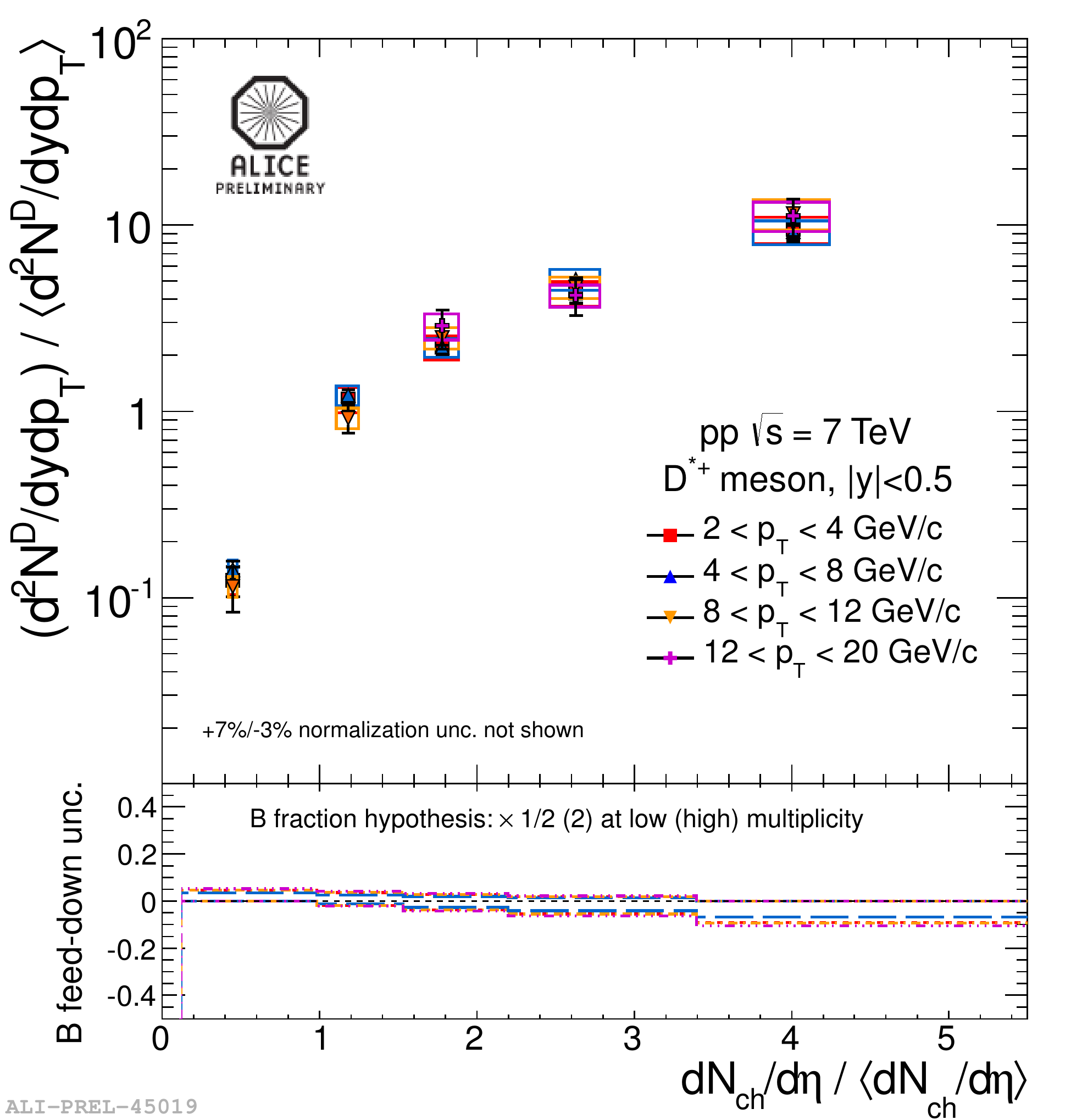}
\end{minipage} 
\caption{\label{fig:Dmesonvsmult} Preliminary results on self-normalized yields of D$^0$, D$^+$ and D$^{*+}$ as a function of
the relative multiplicity of charged particles produced in pp collision at $\sqrt{s}=7$~TeV for several $p_\mathrm{T}$ intervals.}
\end{figure}

The production of D mesons was also studied in several $p_\mathrm{T}$ intervals as a function of the multiplicity of charged particles generated in the collision. The 
multiplicity estimator used in this analysis is the number of SPD tracklets $N_\mathrm{trk}$ (combination of two hits in the two SPD layers) in $|\eta|<1.0$. Since the 
pseudorapidity coverage of the SPD changes with the $z$ position of the interaction vertex, a correction to the measured $N_\mathrm{trk}$ is applied event-by-event. 
The D-meson yields are extracted in five intervals of multiplicity. Figure~\ref{fig:Dmesonvsmult} shows the yield of D$^0$, D$^+$ and D$^{*+}$ in a given multiplicity interval 
divided by the value integrated over multiplicity as a function of relative charged-particle multiplicity (in units of the average) for different $p_\mathrm{T}$ intervals in pp collisions at $\sqrt{s}=7$~TeV. 
The results for D$^0$, D$^+$ and D$^{*+}$ are compatible within uncertainties and show an increase of the D-meson yields with the 
charged-particle multiplicity of the event. A similar trend has also been observed for inclusive J/$\psi$ measured with ALICE~\cite{aliceJpsi}. These results may 
suggest that either D-meson production in pp collisions is connected with a strong hadronic activity, or with multiple hard parton interactions.

\section{Results in Pb--Pb collisions} \label{sec:pbpbresults}
The Pb--Pb data at $\sqrt{s_\mathrm{NN}}=2.76$~TeV were collected in 2010 and 2011, using both minimum-bias and centrality triggers. In 2010, the data were 
collected with a minimum-bias trigger ($L_{\mathrm{int}} = 2.12~\mu\mathrm{b}^{-1}$), while in 2011 an online selection based on the VZERO signal amplitude was 
used to enhance the sample of central (0--10\%, $L_{\mathrm{int}} = 28~\mu\mathrm{b}^{-1}$) and mid-central (10--50\%, $L_{\mathrm{int}} = 6~\mu\mathrm{b}^{-1}$) collisions. In order 
to obtain the pp reference spectrum at $\sqrt{s}=2.76$~TeV, the high-statistics 7~TeV data were scaled to 2.76~TeV using FONLL. The nuclear modification factors 
of D$^0$, D$^+$ and D$^{*+}$ mesons measured in the range $2 < p_\mathrm{T} < 16$~GeV/$c$ using the 2010 data sample were reported in Ref.~\cite{alicepbpbraa} 
and the $v_2$ results from the 2011 data sample were reported in Ref.~\cite{alicepbpbv2,alicepbpbv2Long}.
\begin{figure}[h!t]
\centering
\begin{minipage}[t]{14.8pc}\vspace{0px}
  \includegraphics[width=14.8pc]{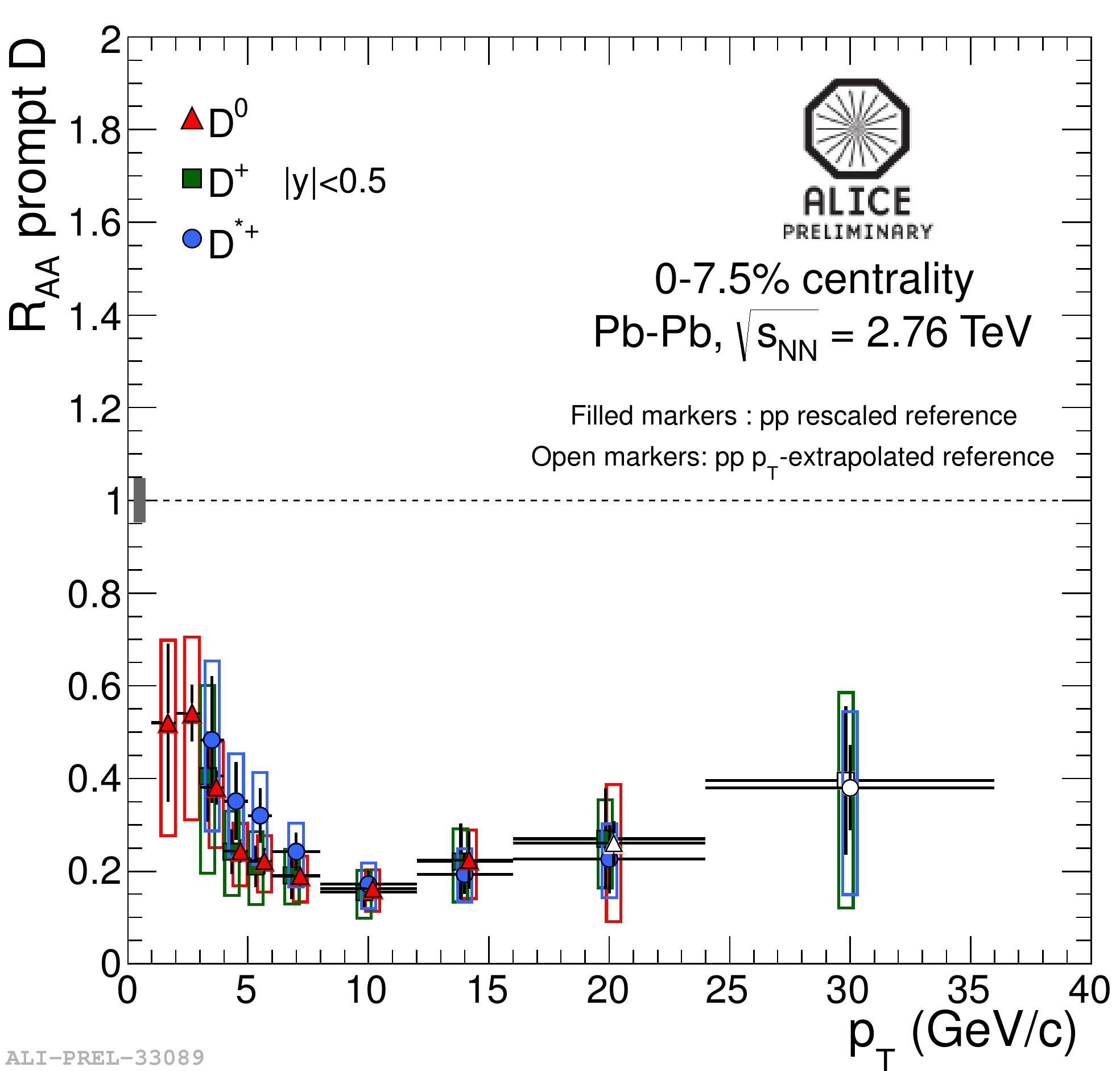}
\end{minipage}\hspace{2pc}%
\begin{minipage}[t]{14.8pc}\vspace{0px}
 \includegraphics[width=14.8pc]{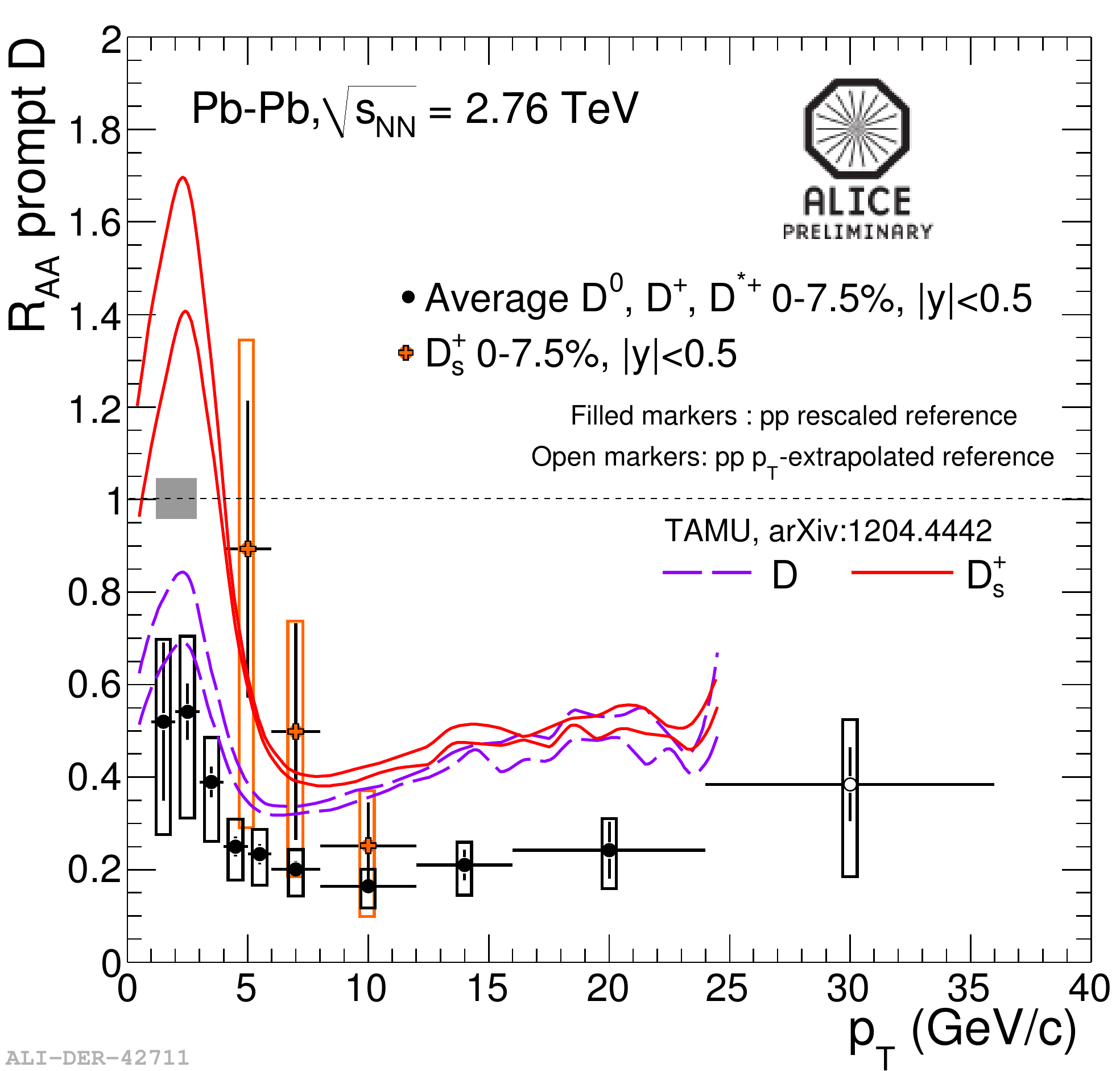}\vspace{0px}
\end{minipage} 
\caption{\label{fig:centralraa} Left panel: $R_\mathrm{AA}$ of D$^0$, D$^+$, and D$^{*+}$ mesons in 
central (0--7.5\%) Pb--Pb collisions at $\sqrt{s_\mathrm{NN}}=2.76$~TeV. Right panel: Average $R_\mathrm{AA}$ of D$^0$, D$^+$, and D$^{*+}$ mesons compared 
with that of D$_\mathrm{s}^+$ mesons. Theoretical model predictions for $R_\mathrm{AA}$ of non-strange D mesons and the D$_\mathrm{s}^+$ meson 
are shown as lines~\cite{charmrecombin1}.}
\end{figure}

The $R_\mathrm{AA}$ of D$^0$, D$^+$, D$^{*+}$ and D$_\mathrm{s}^+$ mesons was measured in most-central (0--7.5\%) Pb--Pb collisions in a wider transverse 
momentum range $1<p_\mathrm{T}<36$~GeV/$c$, using 2011 data. The left panel of Fig.~\ref{fig:centralraa} shows the $R_\mathrm{AA}$ measurements for 
D$^0$, D$^+$, and D$^{*+}$. They are consistent within statistical uncertainties and show a strong suppression (factor of 4--5) for $p_\mathrm{T} > 5$~GeV/$c$.
The right panel of Fig.~\ref{fig:centralraa} shows $R_\mathrm{AA}$ of the D$_\mathrm{s}^+$ meson in the range $4 < p_\mathrm{T} < 12$~GeV/$c$, compared with 
the average $R_\mathrm{AA}$ of non-strange D mesons. A suppression of the D$_\mathrm{s}^+$ meson is observed for 
$8 < p_\mathrm{T} < 12$~GeV/$c$, in agreement within the uncertainties with that of non-strange D mesons in the same $p_\mathrm{T}$ range. Theoretical calculations
predict that D$_\mathrm{s}^+$ meson yield should be less suppressed at low and intermediate $p_\mathrm{T}$ in comparison to non-strange D mesons, if the dominant 
mechanism for D-meson formation is via recombination of charm quarks with light quarks in the medium~\cite{charmrecombin1,charmrecombin2}. One of 
these theoretical predictions for $R_\mathrm{AA}$ of non-strange D mesons and D$_\mathrm{s}^+$ mesons is shown in the right panel of Fig.~\ref{fig:centralraa}. Present data 
hint at the central value of $R_\mathrm{AA}$ of D$_\mathrm{s}^+$ mesons being higher than that of non-strange D mesons at low $p_\mathrm{T}$, however a more precise 
measurement is needed to draw a firm conclusion as they are compatible within current uncertainties. 
\begin{figure}[h!t]
\centering
\begin{minipage}{13.2pc}
\includegraphics[width=13.2pc]{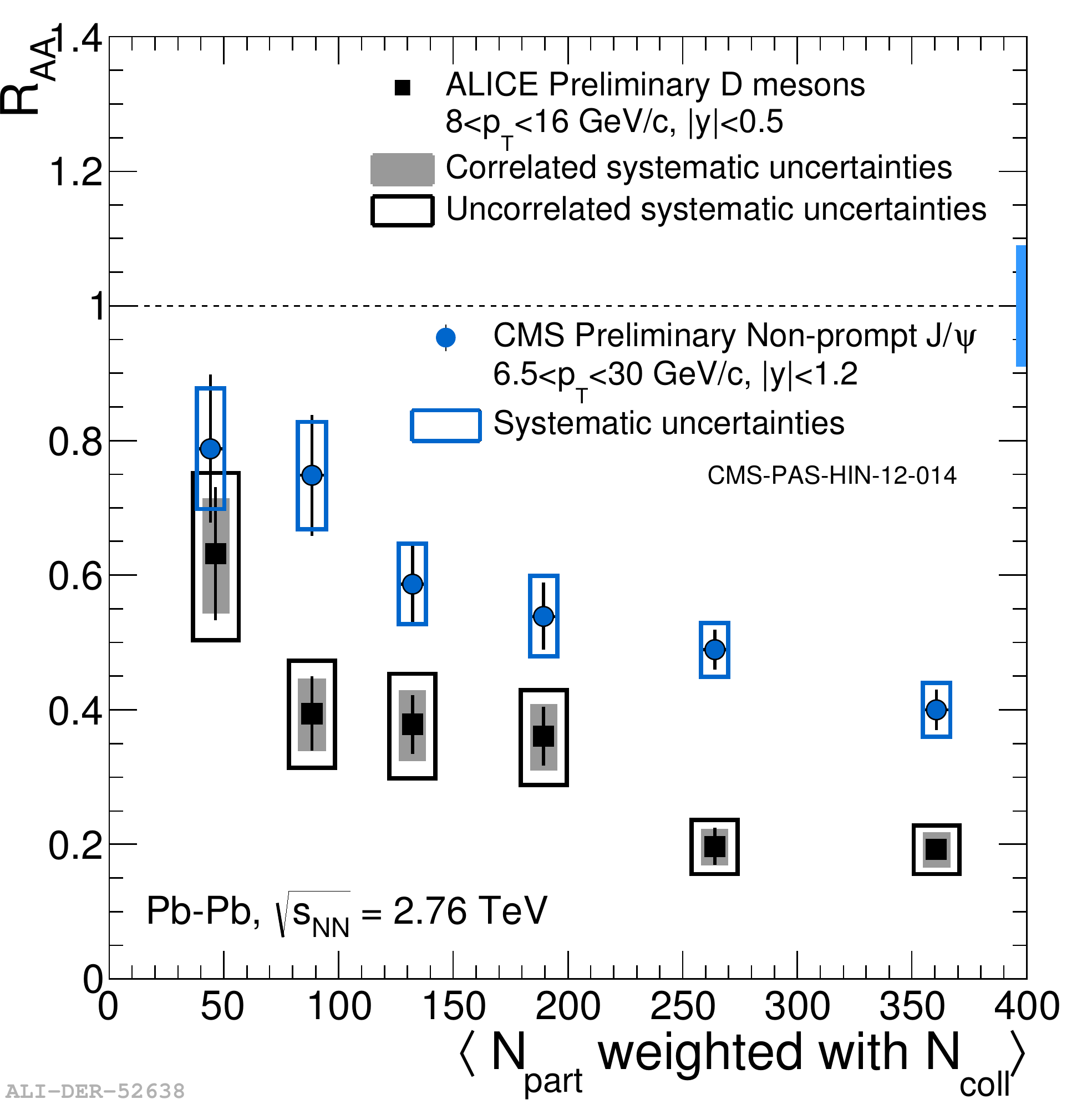}
\end{minipage}\hspace{2pc}%
\begin{minipage}{13.2pc}
\includegraphics[width=13.2pc]{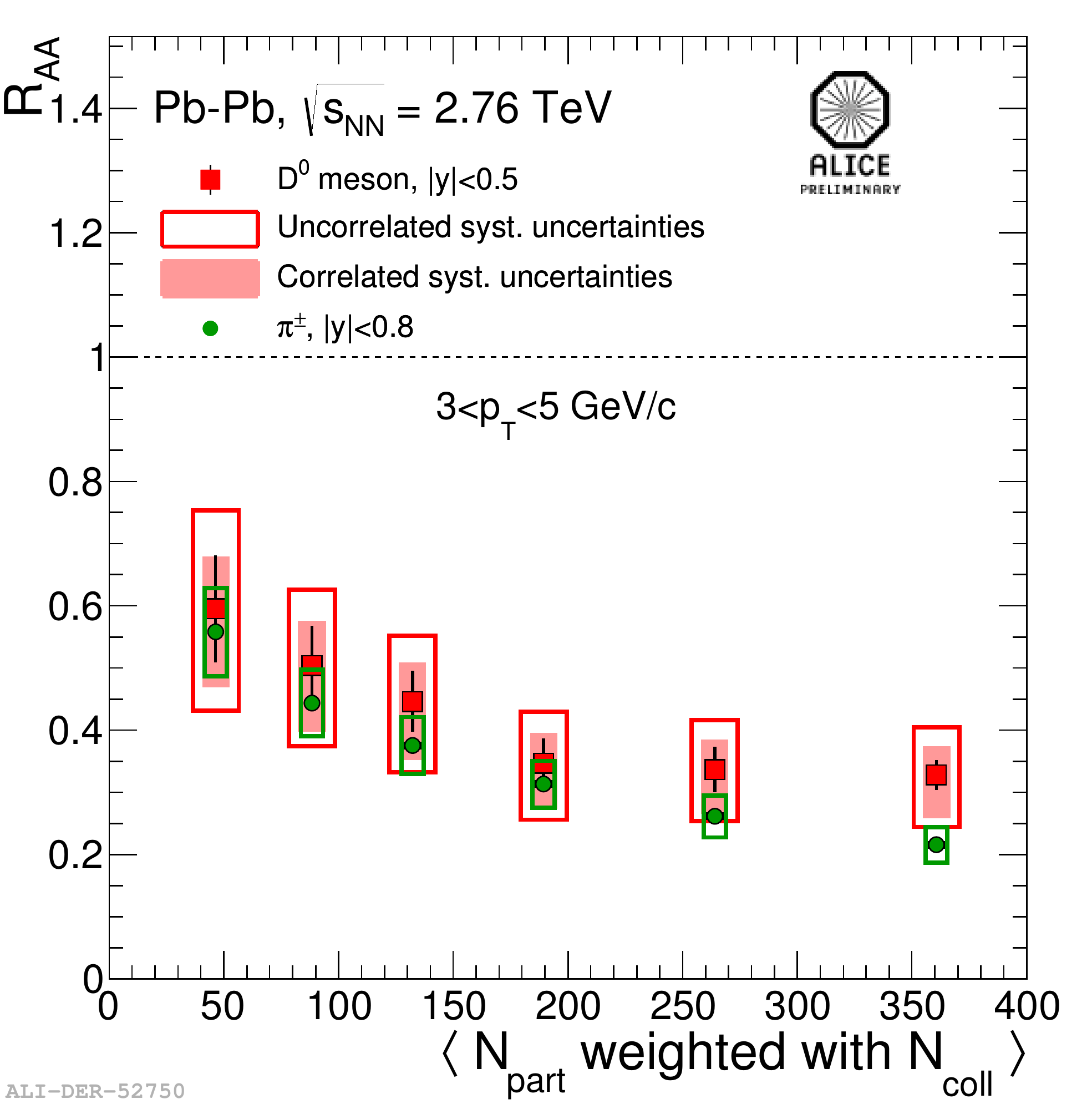}
\end{minipage} 
\caption{\label{fig:nonpromptjpsi} Average $R_\mathrm{AA}$ of D$^0$, D$^+$ and D$^{*+}$ mesons as a function of centrality in Pb--Pb collisions 
at $\sqrt{s_\mathrm{NN}}=2.76$~TeV, compared with $R_\mathrm{AA}$ of non-prompt J/$\psi$ mesons measured with CMS (left panel)~\cite{cmsjpsi}
and charged pions measured with ALICE (right panel)}
\end{figure}

The comparison of the centrality dependence of $R_\mathrm{AA}$ of non-strange D mesons and of J/$\psi$ from B-meson decay measured with CMS~\cite{cmsjpsi} is
shown in the left panel of Fig.~\ref{fig:nonpromptjpsi}. The $p_\mathrm{T}$ ranges for D mesons ($8 < p_\mathrm{T} < 16$~GeV/$c$) and  J/$\psi$ from B-meson 
decays ($6.5 < p_\mathrm{T} < 30$~GeV/$c$) were chosen such that the kinematics of D and B mesons were similar. These results indicate a stronger suppression 
of charm hadrons than beauty hadrons in central Pb–-Pb collisions, as expected from the predicted mass dependence discussed in section~\ref{sec:intro}. The 
right panel of Fig.~\ref{fig:nonpromptjpsi} shows the comparison of 
the centrality dependence of $R_\mathrm{AA}$ of D mesons and charged  pions for $3 < p_\mathrm{T} < 5$~GeV/$c$. The magnitude of the suppression is similar 
for D mesons and pions, which is consistent with some of the calculations discussed in section~\ref{sec:intro}. A hint of a difference in central collisions is 
suggested but more precise measurements are needed to draw a conclusion on the difference between D-meson and pion suppression due to the mass 
and colour-charge dependence of the energy loss.
\begin{figure}[h!t]
\centering
\begin{minipage}{15pc}
\includegraphics[width=15pc]{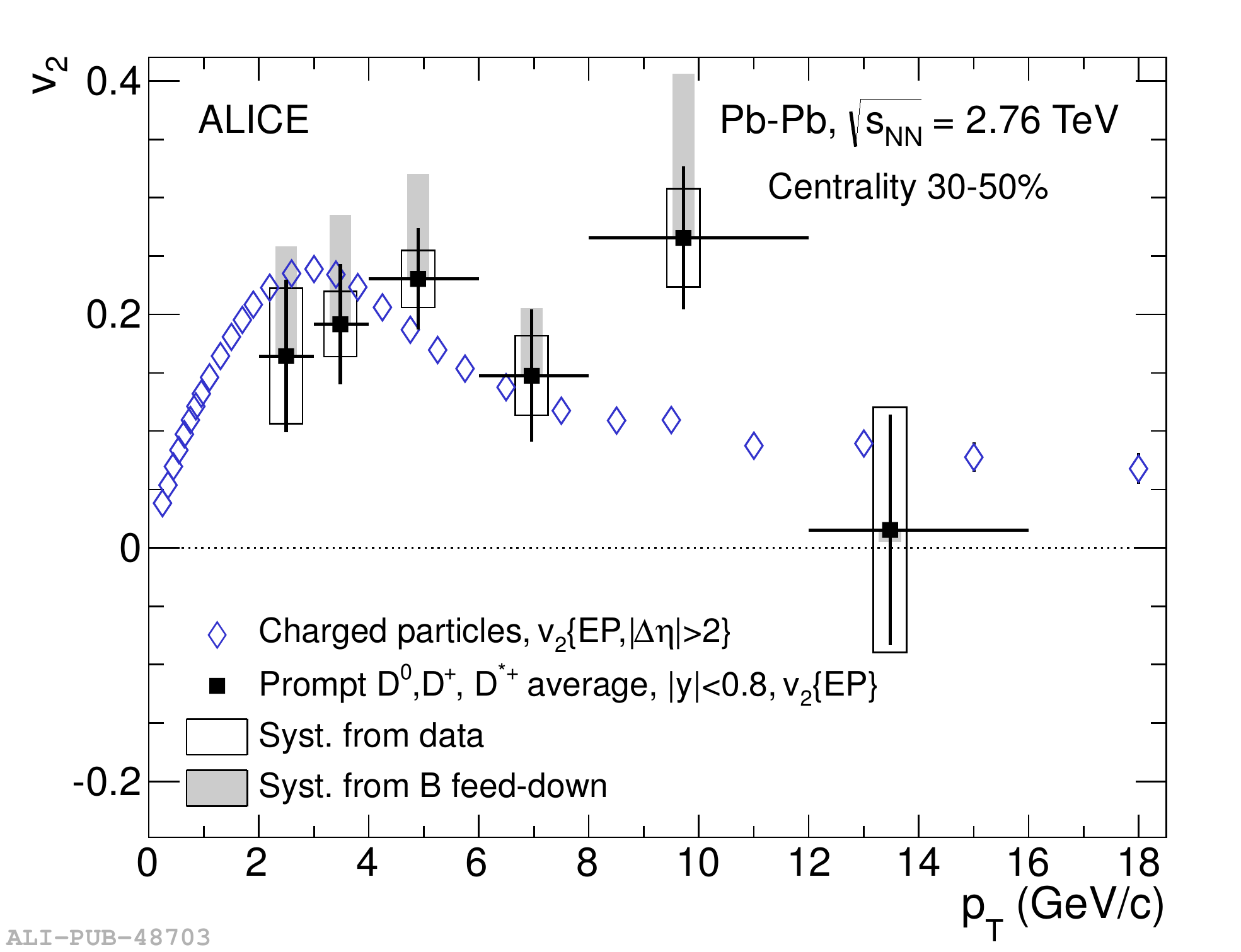}
\end{minipage}\hspace{0pc}%
\begin{minipage}{16pc}
\includegraphics[width=16pc]{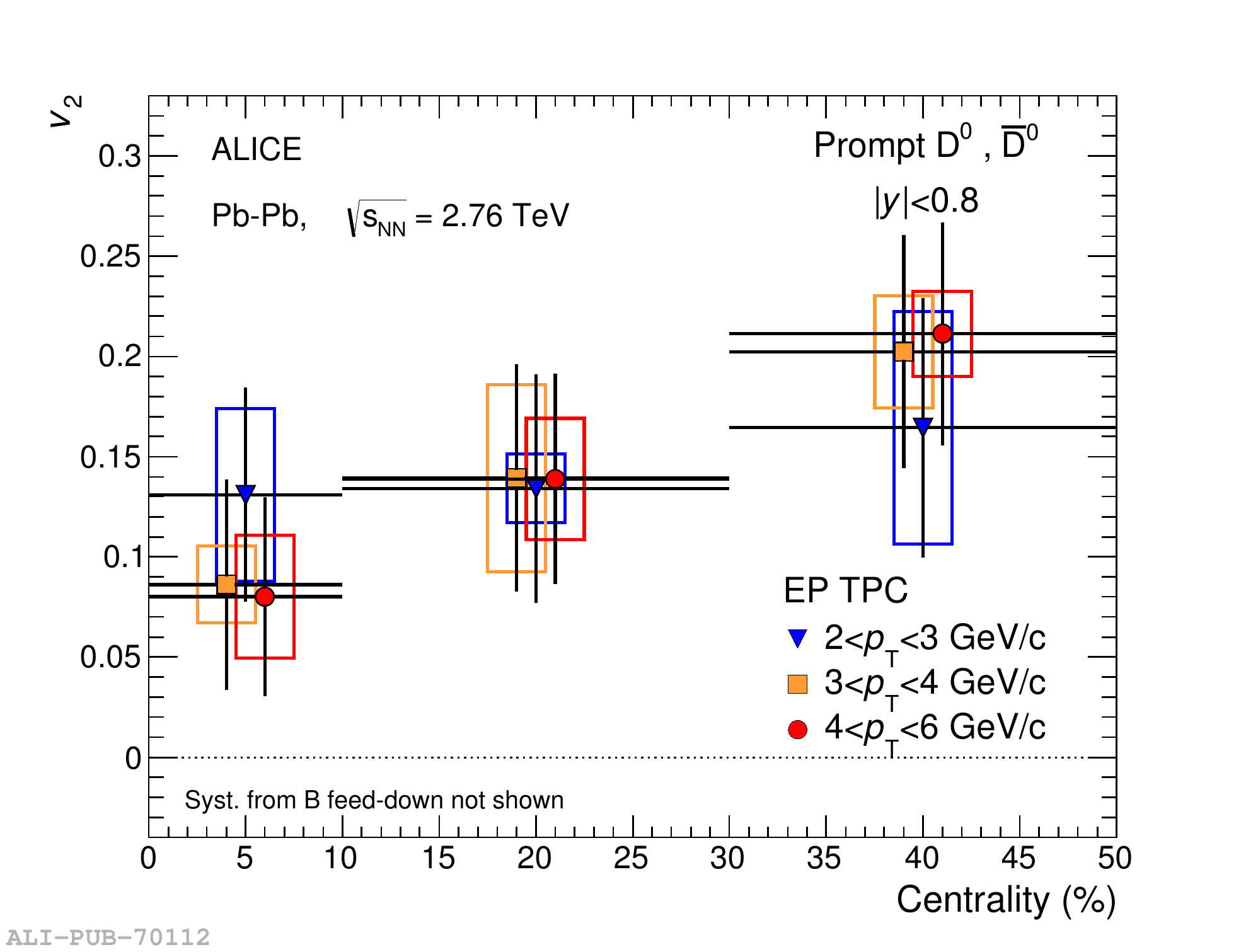}  
\end{minipage} 
\caption{\label{fig:v2daverage} \label{fig:v2vscent} Left panel: Average $v_2$ of D$^0$, D$^+$ and D$^{*+}$ mesons as a function of $p_\mathrm{T}$ for 
Pb--Pb collisions at $\sqrt{s_\mathrm{NN}}=2.76$~TeV in the 30--50\% centrality range, compared with the $v_2$ of charged particles~\cite{alicepbpbv2}. 
Right panel: $v_2$ of D$^0$ in three $p_\mathrm{T}$ intervals as a function of centrality. The data points are displaced horizontally for two of 
the $p_\mathrm{T}$ intervals for better visibility~\cite{alicepbpbv2Long}.}
\end{figure}

The $v_2$ of prompt D mesons at mid-rapidity was measured in the three centrality classes 0--10\%, 10--30\% and 30--50\% using the event-plane 
method~\cite{alicepbpbv2,alicepbpbv2Long}. The average $v_2$ as a function of $p_\mathrm{T}$ for D$^0$, D$^+$ and D$^{*+}$ mesons in the 30--50\% 
centrality class is shown in the left panel of Fig.~\ref{fig:v2daverage}, compared with the measured $v_2$ of charged particles. The D-meson $v_2$ is 
larger than zero with 5$\sigma$ significance in the range $2 < p_\mathrm{T} < 6$~GeV/$c$. A positive $v_2$ is also 
observed for $p_\mathrm{T}>6$~GeV/$c$, which most likely originates from the path-length dependence of the in-medium partonic energy loss, although the large uncertainties 
do not allow a firm conclusion. The measured D-meson $v_2$ is comparable in magnitude with that of charged particles, which are mostly light-flavour hadrons. 
This result suggests that low-momentum charm quarks take part in the collective motion of the system. The centrality dependence of $v_2$ of D$^0$ mesons is 
shown in the right panel of Fig.~\ref{fig:v2daverage} for three  $p_\mathrm{T}$ intervals in the range $2 < p_\mathrm{T} < 6$~GeV/$c$. A decreasing trend of $v_2$ 
towards more central collisions is observed, as expected because of the decreasing initial geometrical anisotropy. 
\begin{figure}[h!t]
\centering
\begin{minipage}{15pc}
\includegraphics[width=15pc]{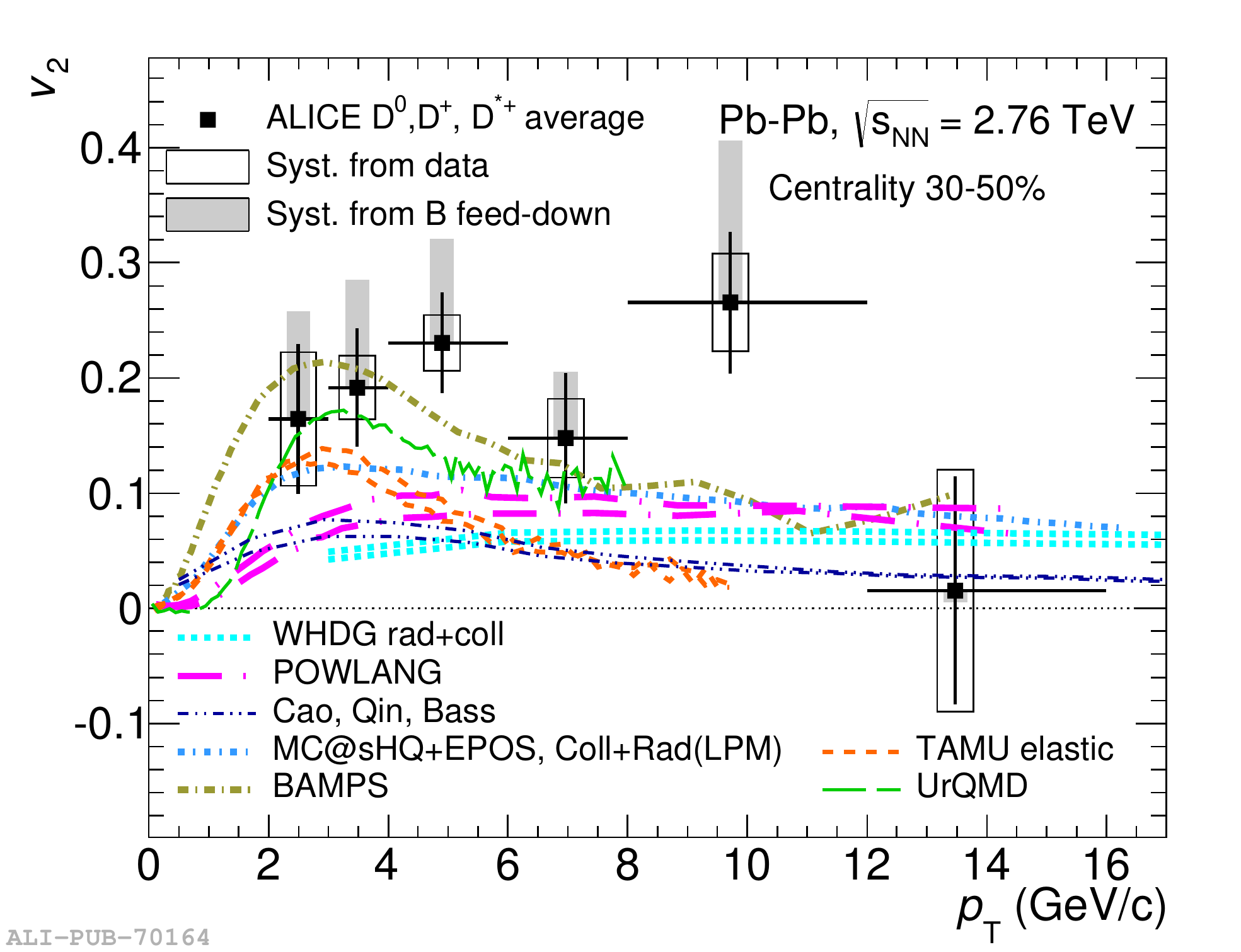}
\end{minipage}\hspace{1.5pc}%
\begin{minipage}{15pc}
\includegraphics[width=15pc]{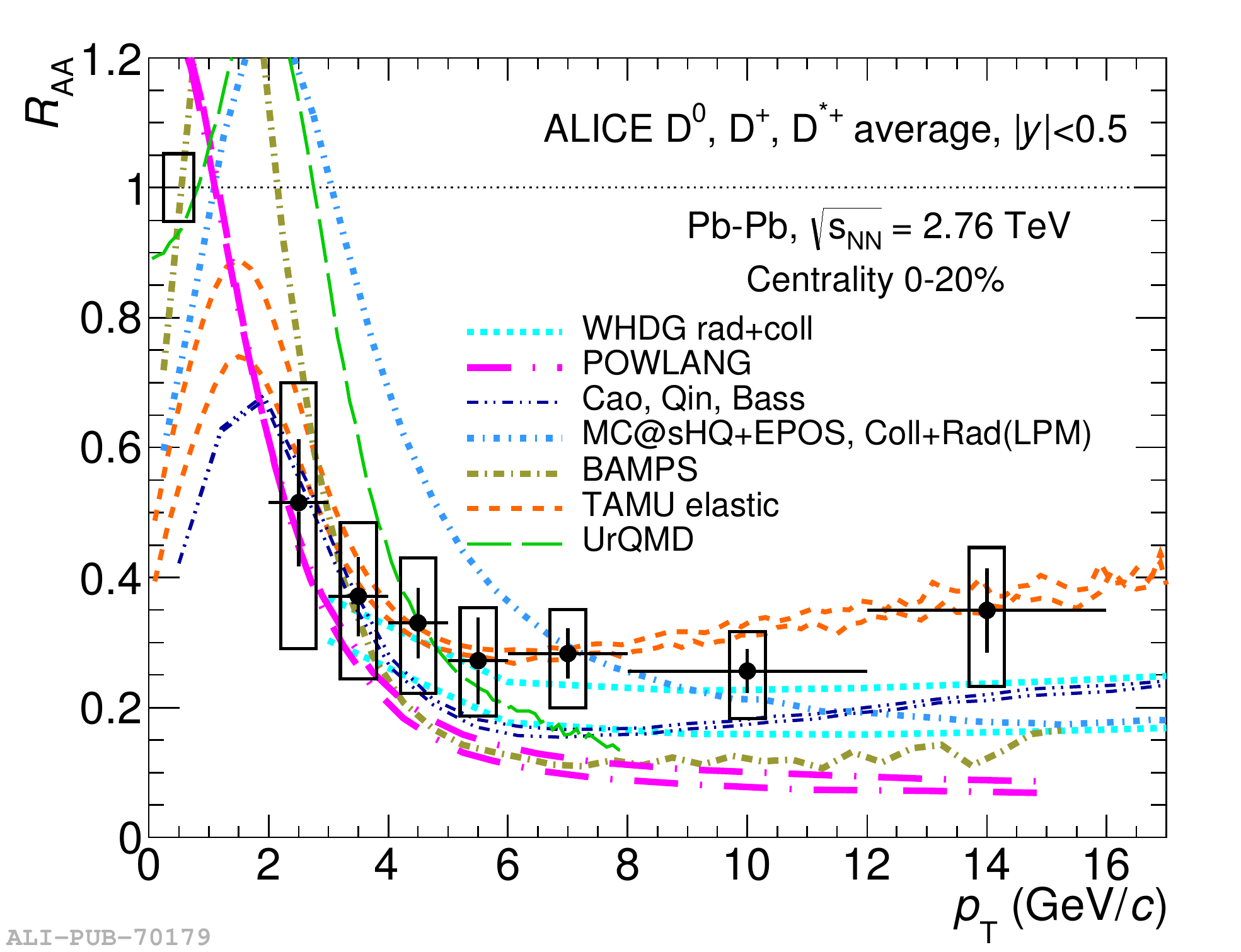}
\end{minipage} 
\caption{\label{fig:RAAv2vsmodels} Left panel: Model comparisons for average $v_2$ of D$^0$, D$^+$ and D$^{*+}$ in the 30--50\% centrality class (left panel), 
average $R_\mathrm{AA}$ of D$^0$, D$^+$ and D$^{*+}$ in the 0--20\% centrality class 
(right panel)~\cite{WHDG,POWLANG,caoqinbass,aichelin1,aichelin2,BAMPS,TAMU,UrQMD1,UrQMD2}.}
\end{figure}

A number of theoretical model calculations are currently available for the elliptic flow coefficient $v_2$  and the nuclear modification factor $R_\mathrm{AA}$ of 
heavy-flavour hadrons~\cite{WHDG,POWLANG,caoqinbass,aichelin1,aichelin2,BAMPS,TAMU,UrQMD1,UrQMD2}. Figure~\ref{fig:RAAv2vsmodels} shows the comparison 
of these models to measurements of the average $v_2$ of D$^0$, D$^+$ and D$^{*+}$ in the 30--50\% centrality class and  average $R_\mathrm{AA}$ of 
D$^0$, D$^+$ and D$^{*+}$ in the 0--20\% centrality class. The anisotropy is qualitatively described by the models that include both charm quark energy loss 
in a geometrically anisotropic medium and mechanisms that transfer the elliptic flow of the medium to charm quarks during the system expansion. The measured 
nuclear modification factor is qualitatively described by the models including in-medium parton energy loss. The model comparison 
for $R_\mathrm{AA}$  and $v_2$ shows that some models describe either the $R_\mathrm{AA}$ or the $v_2$, but a simultaneous description of both observables remains a challenge.

\section{Results in p--Pb collisions}  \label{sec:ppbresults}
\begin{figure}[h!t]
\centering
\begin{minipage}{14pc}
\includegraphics[width=14pc]{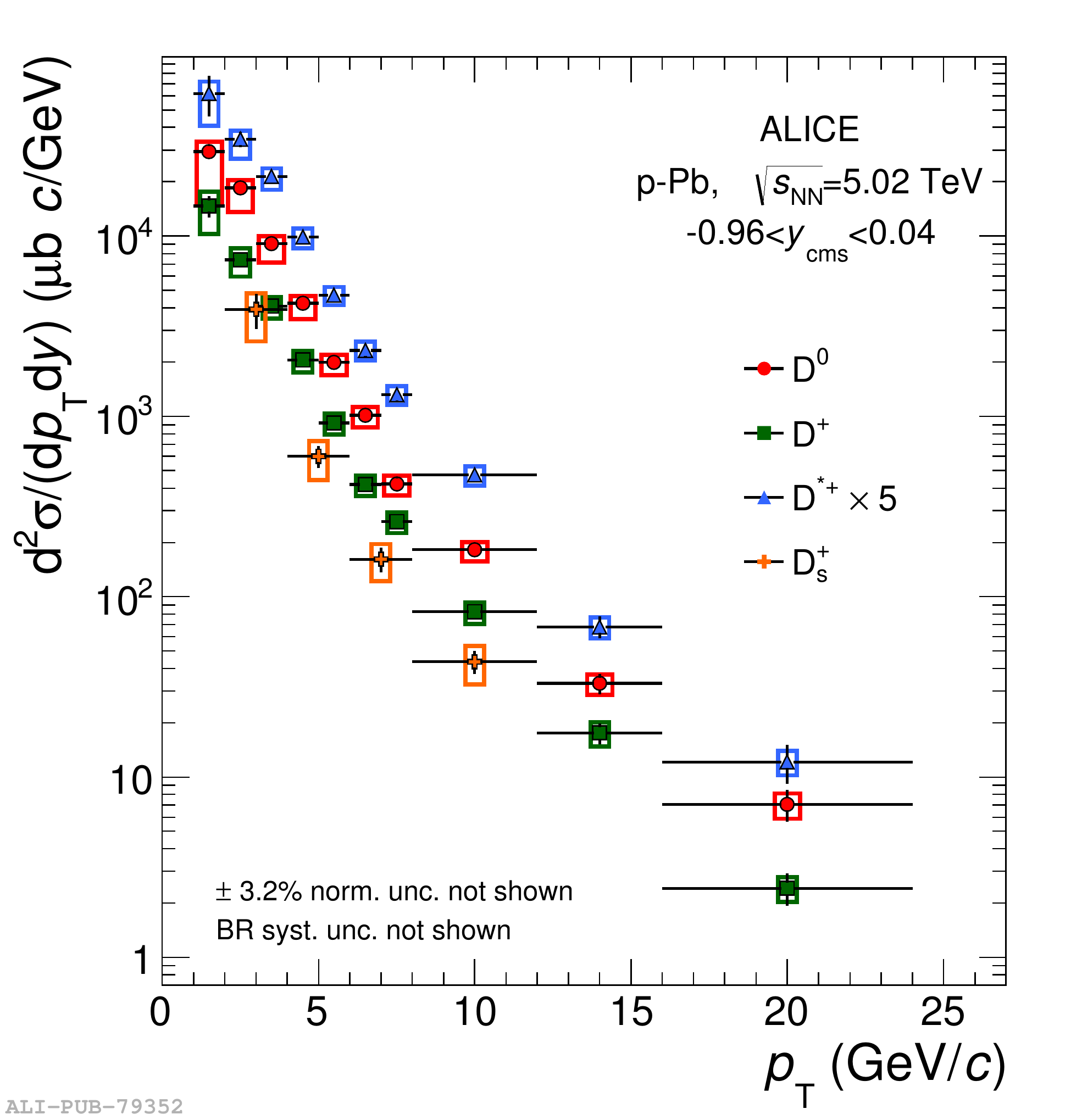}
\end{minipage}\hspace{2pc}%
\begin{minipage}{14pc}
\includegraphics[width=14pc]{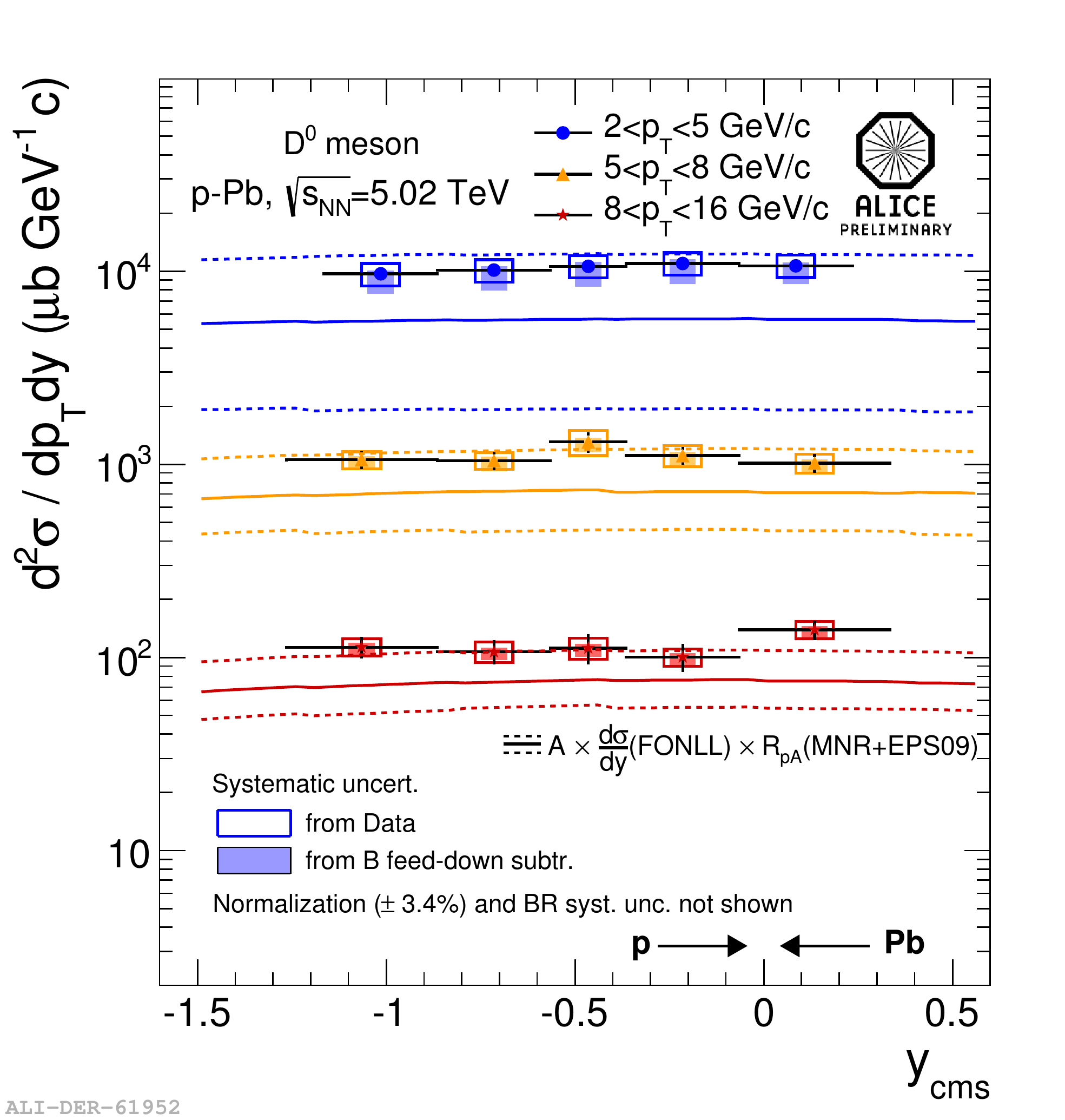}
\end{minipage} 
\caption{\label{fig:Dcrosssection}Left panel: $p_\mathrm{T}$-differential inclusive production cross section of prompt D$^0$, D$^+$, D$^{*+}$, and D$_\mathrm{s}^{+}$ 
mesons in p--Pb collisions at $\sqrt{s_\mathrm{NN}}=5.02$~TeV~\cite{aliceRpPb}. Right panel: $p_\mathrm{T}$, $y$-differential production cross section of 
prompt D$^0$ mesons in p--Pb 
collisions as a function of the rapidity in the centre-of-mass system for three different $p_\mathrm{T}$ ranges. The lines represent expectations 
based on pQCD calculations including the EPS09 parametrization of nuclear PDFs.}
\end{figure}
The p--Pb data at $\sqrt{s_\mathrm{NN}}=5.02$~TeV were collected in 2013 with a minimum-bias trigger. The analysed data sample corresponds to an integrated 
luminosity of $48.6~\mu\mathrm{b}^{-1}$. The D-meson production cross section and the nuclear modification factor $R_\mathrm{pPb}$ were measured in p--Pb collisions 
at $\sqrt{s_\mathrm{NN}}=5.02$~TeV~\cite{aliceRpPb}. The left panel of Fig.~\ref{fig:Dcrosssection} shows the $p_\mathrm{T}$-differential production cross 
sections of prompt D$^0$, D$^+$, D$^{*+}$, and  D$_\mathrm{s}^+$ mesons and the right panel of the same figure shows the production cross section as a function 
of rapidity for D$^0$ in $2 < p_\mathrm{T} < 5$ GeV/$c$, $5 < p_\mathrm{T} < 8$ GeV/$c$ and $8 < p_\mathrm{T} < 16$ GeV/$c$. In the centre of mass system of 
the p--Pb collisions, the measurement covers the rapidity range $-0.96 < y < 0.04$.  Within the current statistical and systematic uncertainties, no evidence of 
a dependence of the production cross section on rapidity is observed within this window. The measured $p_\mathrm{T}$, $y$-differential cross section is compatible 
with predictions obtained by scaling the $y$-differential cross section calculated for pp collisions with FONLL~\cite{FONLL} by the Pb mass number and by a $R_\mathrm{pPb}$ 
estimated, as a function of $p_\mathrm{T}$, on the basis of the MNR~\cite{MNR} calculation and EPS09 nuclear PDF parametrizations~\cite{EPS09}. 
\begin{figure}[h!t]
\centering
\begin{minipage}{14pc}
\includegraphics[width=14pc]{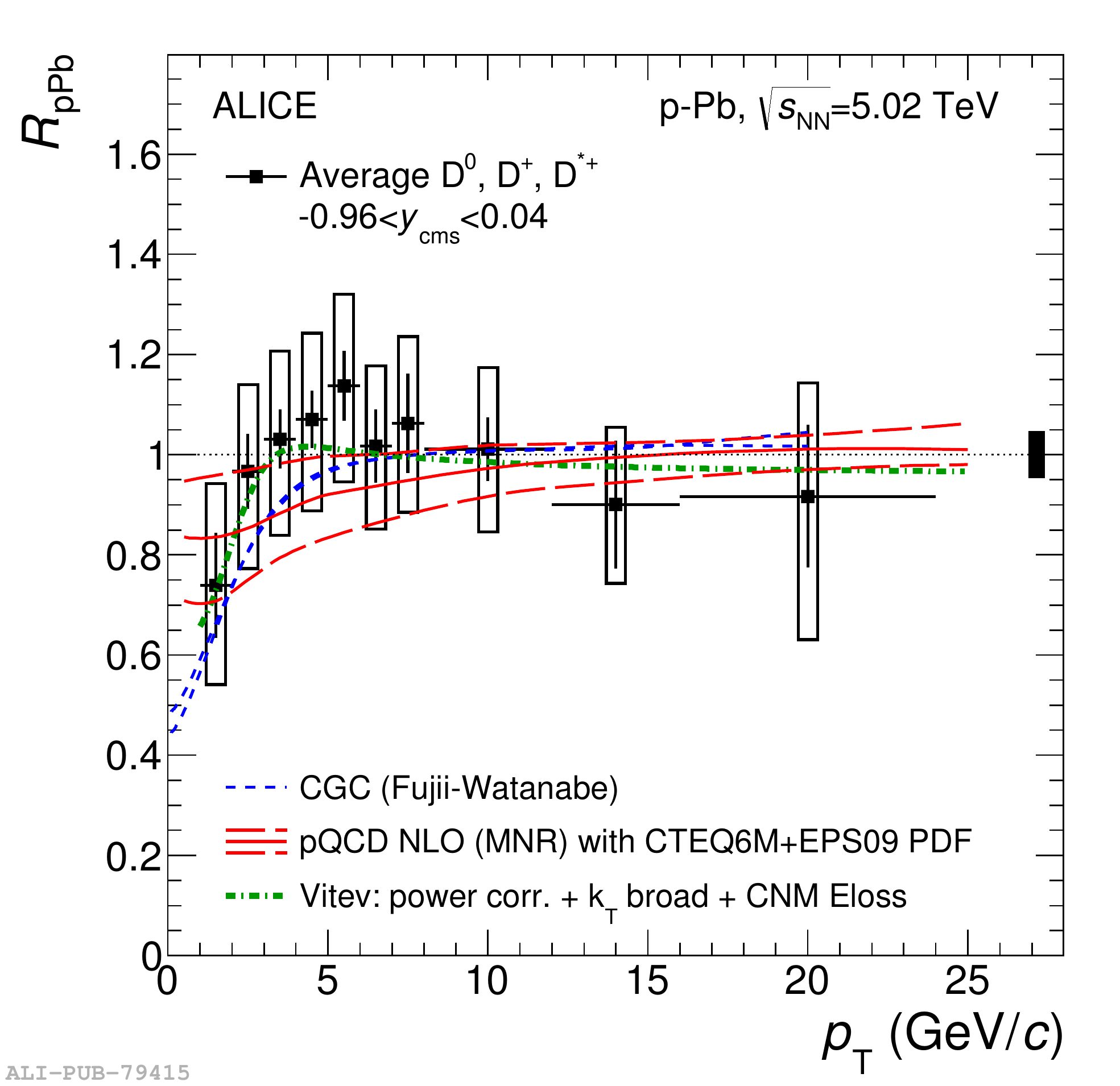}
\end{minipage}\hspace{2pc}%
\begin{minipage}{14pc}
\includegraphics[width=14pc]{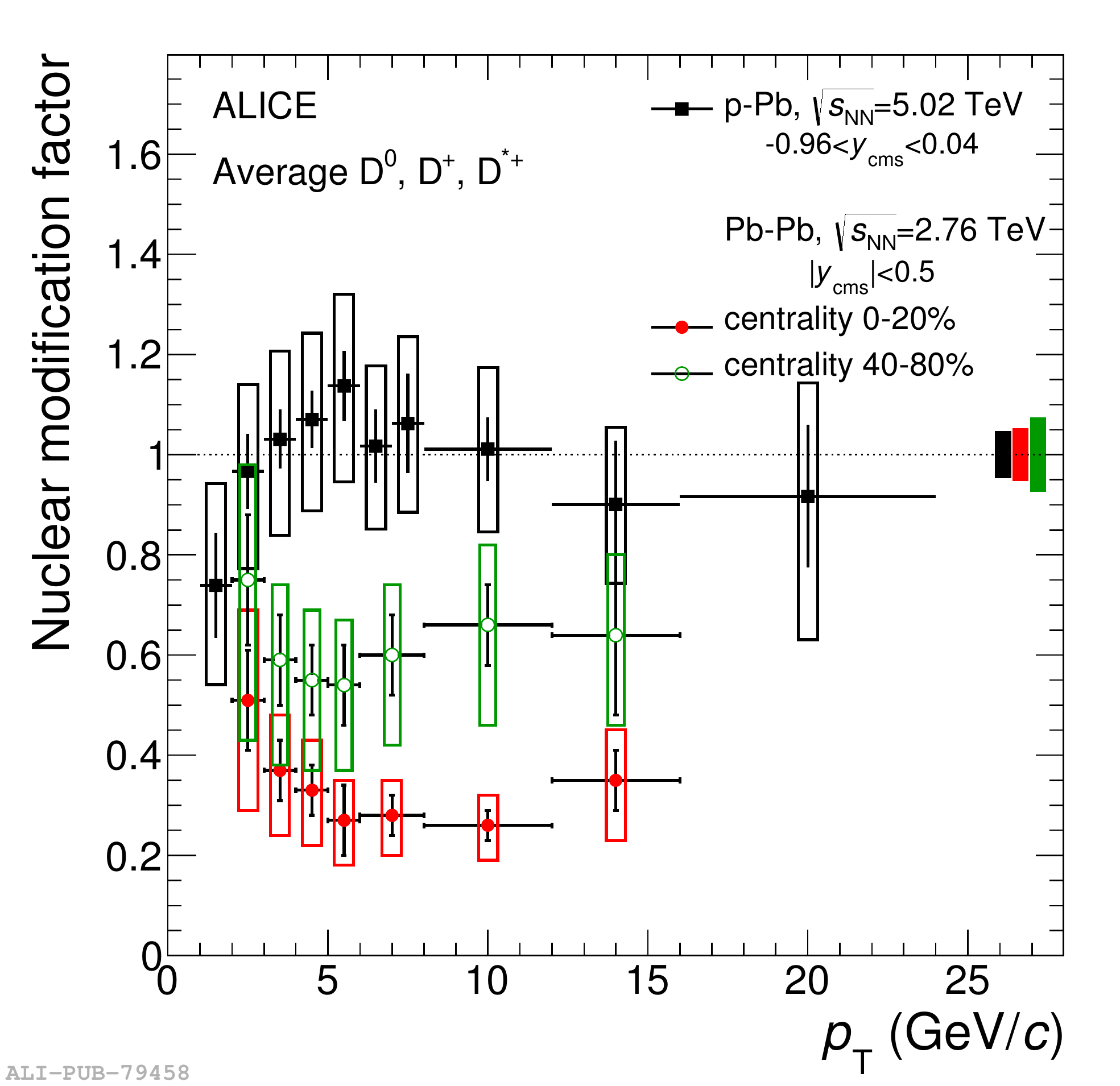}
\end{minipage} 
\caption{\label{fig:rppbmodels}\label{fig:rppbraa} Average $R_\mathrm{pPb}$ of D$^0$, D$^+$ and D$^{*+}$ mesons as a function of $p_\mathrm{T}$ 
at $\sqrt{s_\mathrm{NN}}=5.02$~TeV compared to model calculations~\cite{MNR,EPS09} (left panel), and D-meson $R_\mathrm{AA}$ in central (0--20\%) and 
in semi-peripheral (40--80\%) Pb-–Pb collisions at $\sqrt{s_\mathrm{NN}}=2.76$~TeV  (right panel)~\cite{aliceRpPb}.}
\end{figure}

The $p_\mathrm{T}$-differential cross sections were used to compute the nuclear modification factor $R_\mathrm{pPb}$, defined as 
\begin{equation} \label{eq:rppb}
R_\mathrm{pPb}(p_\mathrm{T}) = \frac{\mathrm{d}\sigma_\mathrm{pPb}/\mathrm{d}p_\mathrm{T}}{A\times(\mathrm{d}\sigma_\mathrm{pp}/\mathrm{d}p_\mathrm{T})},
\end{equation}
where $\sigma_\mathrm{pPb}$ is the cross section in p--Pb collisions, $\sigma_\mathrm{pp}$ is the cross section in pp collisions at the same centre-of-mass energy, and 
$A$ is the mass number of the Pb nucleus. The reference pp cross sections at $\sqrt{s}=5.02$~TeV were obtained by a FONLL based energy scaling of
the $p_\mathrm{T}$-differential cross sections measured at $\sqrt{s}=7$~TeV~\cite{energyscaling}.

The $R_\mathrm{pPb}$ measurements for all D-meson species, including D$_\mathrm{s}^+$, are consistent, and compatible with unity within the uncertainties 
in the measured $p_\mathrm{T}$ range~\cite{aliceRpPb}. The average $R_\mathrm{pPb}$ of D$^0$, D$^+$ and D$^{*+}$ was calculated using the relative statistical 
uncertainties as weights. The left panel of Fig.~\ref{fig:rppbmodels} shows the average $R_\mathrm{pPb}$ of D mesons as a function of 
$p_\mathrm{T}$, compared to model calculations. It can be seen from the present measurement that the cold nuclear matter effects are smaller than 
the uncertainties for $p_{\mathrm{T}} \gtrsim 3$~GeV/$c$. Predictions based on NLO pQCD calculations (MNR~\cite{MNR}) of D-meson production, including 
the EPS09~\cite{EPS09} nuclear modification of the CTEQ6M PDF~\cite{CTEQ6M} and calculations based on the Color Glass Condensate~\cite{CGC} can describe the 
measurement considering only initial state effects. Data are also well described by a calculation including energy loss in cold nuclear matter, nuclear shadowing 
and $k_\mathrm{T}$-broadening~\cite{kTbroad,kTbroad2}. Within uncertainties, the experimental results do not favour one prediction over the other.

In the right panel of Fig.~\ref{fig:rppbraa}, the average $R_\mathrm{AA}$ of prompt D mesons in central (0--20\%) and in semi-peripheral (40--80\%) 
Pb--Pb collisions at $\sqrt{s_\mathrm{NN}}=2.76$~TeV~\cite{alicepbpbraa} is shown along with the average $R_\mathrm{pPb}$ of prompt D mesons in p--Pb collisions at 
$\sqrt{s_\mathrm{NN}}=5.02$~TeV. Although the 
centre-of-mass energies of the two systems differ, there is a clear suppression in Pb--Pb collisions compared with p--Pb for $p_\mathrm{T} \gtrsim 4$~GeV/$c$. 
This indicates that final-state effects, such as partonic energy loss in the medium, play a far more dominant role in the suppression of D-meson production 
at higher $p_\mathrm{T}$ than initial-state effects, such as nuclear shadowing.

\section{Conclusions and Outlook}
Measurements of the production of D mesons at mid-rapidity in pp, Pb--Pb and p--Pb collisions performed with ALICE have been presented. The $p_\mathrm{T}$-differential 
cross sections of D mesons in pp collisions are described within uncertainties by theoretical predictions based on perturbative QCD, such as FONLL, GM-VFNS 
and $k_\mathrm{T}$-factorization at LO. D-meson production was studied as a function of the charged particle multiplicity in pp collisions at $\sqrt{s}=7$~TeV. An increase 
of the D-meson yield as a function of multiplicity is observed, which suggests a connection of D-meson production to a strong hadronic activity or to multiple hard parton 
interactions.

A strong suppression of prompt D mesons is observed at high $p_\mathrm{T}$ in central Pb-Pb collisions at $\sqrt{s_\mathrm{NN}}=2.76$~TeV. The suppression 
is of similar magnitude for all non-strange D mesons. Data are not conclusive on a possible enhancement of the D$_\mathrm{s}^+$ yield relative to non-strange D mesons at 
low $p_\mathrm{T}$. The observed difference in the suppression of D mesons and non-prompt J/$\psi$ from B-meson decays measured with CMS at high $p_\mathrm{T}$ in 
central collisions is consistent with predictions of a mass hierarchy in the energy loss of heavy quarks. 

A $v_2$ significantly larger than zero was measured for D mesons at low $p_\mathrm{T}$ in semi-peripheral Pb-Pb collisions, suggesting that charm quarks take part 
in the collective expansion of the medium. The measured $R_\mathrm{AA}$ and $v_2$ of D mesons were compared with
several theoretical model calculations. Although some models can describe measurements of $R_\mathrm{AA}$  and $v_2$ separately, it is challenging for models to 
describe simultaneously the large suppression of D mesons in central collisions and their azimuthal anisotropy in non-central collisions. $R_\mathrm{pPb}$ for 
D mesons as a function of $p_\mathrm{T}$ in p--Pb collisions is consistent with unity and also compatible with models including initial-state effects. This suggests that 
the suppression visible at high $p_\mathrm{T}$ in Pb--Pb collisions primarily is due to the final-state effects rather than initial-state effects.

After a shutdown of close to 2 years, the LHC will resume operations in 2015 with an energy almost a factor of two higher than in the previous running period
and it is expected to deliver a factor 5--10 larger integrated luminosity for Pb--Pb collisions prior to the second long shutdown. The increased statistics resulting 
from this improvement will improve the precision of D-meson measurements with ALICE and shed light on the intriguing questions opened by the present data.

\end{document}